\documentclass[
preprint,
superscriptaddress,
preprintnumbers,
nofootinbib,
showkeys,
amsmath, amssymb,
aps,
pra,
]{revtex4-2}
\usepackage[utf8]{inputenc}
\usepackage[T1]{fontenc}
\usepackage{lipsum}

\usepackage[normalem]{ulem}

\usepackage[english]{babel}
\usepackage{amsmath, amssymb, bm}
\usepackage{amsfonts}  
\usepackage{mathtools} 
\usepackage[normalem]{ulem} 

\usepackage[hang]{caption}
\usepackage{graphicx, subcaption} 
\usepackage{tabularx}
\usepackage{setspace} 

\usepackage{ifthen} 
\usepackage[ligature]{semantic} 
\usepackage{siunitx} 

\usepackage{color}
\usepackage{transparent}	

\definecolor{darkgreen}{RGB}{0,100,0}

\newcommand{\AdS}{\mathrm{AdS}}

\newcommand{\SNG}{\ensuremath{S_\mathrm{N.G.}^{}}}
\newcommand{\zmin}{\ensuremath{z_{\mathrm{min}}}}

\newcommand{\zminbar}{\ensuremath{\bar{z}_\mathrm{min}}}

\newcommand{\bvec}[1]{\ensuremath{\bm{#1}}}
\let\mgt\gg 
\let\mlt\ll 

\renewcommand{\gg}{\bvec{g}}

\renewcommand{\ll}{\bvec{\ell}}

\newcommand  {\xx}{\bvec{x}}

\newlength\bracketsize

\newcommand{\PAR}[2][-1]{
    \setlength{\bracketsize}{0.75ex * #1 + 0.50ex}
    \ifthenelse
    { \equal{#1}{-1} }
    { \ensuremath{ {\left( #2 \right) } } }
    { \ensuremath{
            {               \left( \rule{0pt}{ \bracketsize } \right. \hspace{-1pt} }
            #2
            { \hspace{-1pt} \left. \rule{0pt}{ \bracketsize } \right)               }
} } }

\newcommand{\COL}[2][-1]{
    \setlength{\bracketsize}{0.75ex * #1 + 0.50ex}
    \ifthenelse
    { \equal{#1}{-1} }
    { \ensuremath{ {\left[\, #2 \,\right] } } }
    { \ensuremath{
            {               \left[ \rule{0pt}{ \bracketsize } \right. \hspace{-1pt} }
            \,#2\,
            { \hspace{-1pt} \left. \rule{0pt}{ \bracketsize } \right]               }
} } }

\newcommand{\CHA}[2][-1]{
    \setlength{\bracketsize}{0.75ex * #1 + 0.50ex}
    \ifthenelse
    { \equal{#1}{-1} }
    { \ensuremath{ {\left\{ #2 \right\} } } }
    { \ensuremath{
            {              \left\{ \rule{0pt}{ \bracketsize} \right.  \hspace{-1pt} }
            #2
            { \hspace{-1pt} \left. \rule{0pt}at{ \bracketsize} \right\}               }
} } }

\newcommand{\at}[2][-1]{
    \setlength{\bracketsize}{0.75ex * #1 + 0.50ex}
    \ifthenelse
    { \equal{#1}{-1} }
    { \ensuremath{ {\left.\, #2 \right| } } }
    { \ensuremath{
            {           \!\!\left. \rule{0pt}{ \bracketsize} \right. \hspace{-1pt} }
            \,#2
            { \hspace{-1pt} \left. \rule{0pt}{ \bracketsize} \right|              }
} } }

\newcommand{\abs}[2][-1]{
    \setlength{\bracketsize}{0.75ex * #1 + 0.50ex}
    \ifthenelse
    { \equal{#1}{-1} }
    { \ensuremath{ {\left| #2 \right| } } }
    { \ensuremath{
            {               \left| \rule{0pt}{ \bracketsize} \right. \hspace{-1pt} }
            #2
            { \hspace{-1pt} \left. \rule{0pt}{ \bracketsize} \right|               }
} } }

\newcommand{\average}[2][0]{
    \setlength{\bracketsize}{0.75ex * #1 + 0.50ex}
    \ifthenelse
    { \equal{#1}{-1} }
    { \ensuremath{ {\left< #2 \right> } } }
    { \ensuremath{
        {               \left< \rule{0pt}{ \bracketsize} \right. \hspace{-1pt} }
        #2
        { \hspace{-1pt} \left. \rule{0pt}{ \bracketsize} \right>               }
} } }

\newcommand{\complexi}{\hspace{.15ex}\textrm{i}\hspace{.15ex}}
\newcommand{\exponentiale}{\hspace{.15ex}\textrm{e}\hspace{.15ex}}

\makeatletter

\@namedef{u8:\detokenize{α}}{\alpha}
\@namedef{u8:\detokenize{β}}{\beta}
\@namedef{u8:\detokenize{γ}}{\gamma}
\@namedef{u8:\detokenize{δ}}{\delta}
\@namedef{u8:\detokenize{ϵ}}{\epsilon}
\@namedef{u8:\detokenize{ε}}{\varepsilon}
\@namedef{u8:\detokenize{ζ}}{\zeta}
\@namedef{u8:\detokenize{η}}{\eta}
\@namedef{u8:\detokenize{θ}}{\theta}
\@namedef{u8:\detokenize{ϑ}}{\vartheta}
\@namedef{u8:\detokenize{κ}}{\kappa}
\@namedef{u8:\detokenize{λ}}{\lambda}
\@namedef{u8:\detokenize{μ}}{\mu}
\@namedef{u8:\detokenize{ν}}{\nu}
\@namedef{u8:\detokenize{ξ}}{\xi}
\@namedef{u8:\detokenize{π}}{\pi}
\@namedef{u8:\detokenize{ρ}}{\rho}
\@namedef{u8:\detokenize{ϱ}}{\varrho}
\@namedef{u8:\detokenize{σ}}{\sigma}
\@namedef{u8:\detokenize{ς}}{\varsigma}
\@namedef{u8:\detokenize{τ}}{\tau}
\@namedef{u8:\detokenize{υ}}{\upsilon}
\@namedef{u8:\detokenize{ϕ}}{\phi}
\@namedef{u8:\detokenize{φ}}{\varphi}
\@namedef{u8:\detokenize{χ}}{\chi}
\@namedef{u8:\detokenize{ψ}}{\psi}
\@namedef{u8:\detokenize{ω}}{\omega}
\@namedef{u8:\detokenize{ϖ}}{\varpi}

\@namedef{u8:\detokenize{Γ}}{\Gamma}
\@namedef{u8:\detokenize{Δ}}{\Delta}
\@namedef{u8:\detokenize{Θ}}{\Theta}
\@namedef{u8:\detokenize{Λ}}{\Lambda}
\@namedef{u8:\detokenize{Ξ}}{\Xi}
\@namedef{u8:\detokenize{Π}}{\Pi}
\@namedef{u8:\detokenize{Σ}}{\Sigma}
\@namedef{u8:\detokenize{ϒ}}{\Upsilon}
\@namedef{u8:\detokenize{Φ}}{\Phi}
\@namedef{u8:\detokenize{Ψ}}{\Psi}
\@namedef{u8:\detokenize{Ω}}{\Omega}

\@namedef{u8:\detokenize{±}}{\pm}
\@namedef{u8:\detokenize{²}}{^{2}}
\@namedef{u8:\detokenize{³}}{^{3}}
\@namedef{u8:\detokenize{¹}}{^{1}}
\@namedef{u8:\detokenize{⁰}}{^{0}}
\@namedef{u8:\detokenize{·}}{\hspace{-0.05ex}\cdot\hspace{-0.05ex}}
\@namedef{u8:\detokenize{×}}{\hspace{-0.05ex}\times\hspace{-0.05ex}}
\@namedef{u8:\detokenize{î}}{\complexi}
\@namedef{u8:\detokenize{ĩ}}{\complexi}
\@namedef{u8:\detokenize{ħ}}{\hbar}
\@namedef{u8:\detokenize{ḏ}}{\mathrm{d}}
\@namedef{u8:\detokenize{ē}}{\exponentiale}
\@namedef{u8:\detokenize{†}}{\dagger}
\@namedef{u8:\detokenize{‡}}{\ddagger}
\@namedef{u8:\detokenize{ı}}{\imath}
\@namedef{u8:\detokenize{ℓ}}{\ell}
\@namedef{u8:\detokenize{←}}{\leftarrow}
\@namedef{u8:\detokenize{→}}{\rightarrow}
\@namedef{u8:\detokenize{∂}}{\partial}
\@namedef{u8:\detokenize{≠}}{\neq}
\@namedef{u8:\detokenize{≤}}{\leq}
\@namedef{u8:\detokenize{≥}}{\geq}
\@namedef{u8:\detokenize{∞}}{\infty}
\@namedef{u8:\detokenize{∫}}{\int}
\@namedef{u8:\detokenize{≈}}{\approx}
\@namedef{u8:\detokenize{≡}}{\equiv}
\@namedef{u8:\detokenize{□}}{\Box}
\@namedef{u8:\detokenize{ⵈ}}{\cdots}
\@namedef{u8:\detokenize{ø}}{\phantom}

\makeatother

\expandafter\newcommand\csname u8:\detokenize{√}\endcsname[2][]{%
    \ifthenelse
    { \equal{#1}{} }
    { \sqrt{#2} }
    { \sqrt[#1]{#2} }
}

\mathlig{<<}{\mlt}
\mathlig{>>}{\mgt}
\mathlig{!=}{\neq}
\mathlig{=!}{\neq}
\mathlig{/=}{\neq}
\mathlig{=/}{\neq}
\mathlig{<>}{\neq}
\mathlig{><}{\neq}
\mathlig{==}{\equiv}
\mathlig{='}{\simeq}
\mathlig{<=}{\leq}
\mathlig{=<}{\leq}
\mathlig{>=}{\geq}
\mathlig{=>}{\geq}

\mathlig{+-}{\pm}
\mathlig{-+}{\mp}
\mathlig{-->}{\rightarrow}
\mathlig{--->}{\longrightarrow}
\mathlig{<--}{\leftarrow}
\mathlig{<---}{\longleftarrow}
\mathlig{<->}{\leftrightarrow}
\mathlig{<-->}{\longleftrightarrow}
\mathlig{|->}{\mapsto}
\mathlig{|-->}{\longmapsto}
\mathlig{==>}{\Rightarrow}
\mathlig{===>}{\Longrightarrow}
\mathlig{==>}{\Leftarrow}
\mathlig{===>}{\Longleftarrow}
\mathlig{<=>}{\leftrightarrow}
\mathlig{<==>}{\longleftrightarrow}

\begin{document}


\title{Holography and charmonium structure in a finite density plasma}

\author{Nelson R. F. Braga}
\email{braga@if.ufrj.br}
\affiliation{Instituto de Física, Universidade Federal do Rio de Janeiro, Caixa Postal 68528, RJ 21941-972, Brazil.}

\author{William S. Cunha}
\email{wscunha@pos.if.ufrj.br}
\affiliation{Instituto de Física, Universidade Federal do Rio de Janeiro, Caixa Postal 68528, RJ 21941-972, Brazil.}

\author{Yan F. Ferreira}
\email{yancarloff@pos.if.ufrj.br}
\affiliation{Instituto de Física, Universidade Federal do Rio de Janeiro, Caixa Postal 68528, RJ 21941-972, Brazil.}

\begin{abstract}
 It has recently been proposed that the extra dimension in holographic models for charmoniun is related to its internal structure. Representing the interaction between the quark anti-quark pair by a string inside the background used in these models, the linear term of the Cornell potential was obtained. More than that, the dissociation in the thermal medium is also described in a consistent way. Here we extend this study to the case of a plasma with finite density. The combined effects of density and temperature are analyzed from the point of view of quark anti-quark interaction and the results obtained are consistent with the ones derived previously using spectral functions.

\end{abstract}


\maketitle
\newpage

\vspace{.75\baselineskip}
\section{Introduction}

The quark-gluon plasma (QGP) has been a subject of great interest for nuclear and particle physicists since it was indirectly observed. This very short-lived state of matter, formed in ultra relativistic heavy ion collisions, behaves like a perfect fluid, having the lowest viscosity ever measured \cite{Shuryak:2008eq,Casalderrey-Solana:2011dxg}. It is formed at temperatures of the order of hundreds of $\SI{}{\mega\electronvolt}$. 
The collisions initially produce a huge amount of hadrons that then dissociate, when the plasma is formed. However, heavy mesons like charmonium and bottomonium may survive the dissociation process, depending on the temperature, density, and other properties of the plasma. Charmonium suppression is interpreted as an evidence of formation of the QGP \cite{Matsui:1986dk,Satz:2005hx} and the intensity of such a process is related to the properties of the plasma. This is the main reason why it is important to understand how heavy mesons behave in a thermal medium.

Holographic models have been used successfully to describe the dissociation of mesons in the QGP. In particular, the tangent model proposed in \cite{Braga:2017bml} contains three energy parameters, related to the heavy quark mass, the intensity of the strong interaction (string tension) and the scale of energy change in the non-hadronic decay of quarkonium. It was used for fitting masses and decay constants of charmonium and bottomonium states. Then it was applied to calculate the spectral functions of heavy mesons in a finite temperature plasma with density, magnetic moment, and angular momentum \cite{Braga:2017oqw, Braga:2018zlu, Braga:2018hjt, Braga:2019yeh, Braga:2020hhs, Braga:2023fac}. 

In all these works, inspired by the AdS/CFT correspondence, the vector meson quasi-states are represented by a field living in a five dimensional background. In the recent work of Ref.~\cite{Braga:2024heu}, it was shown that one can consider a string in the same background as representing the interaction energy between the quark and anti-quark that form a charmonium state. Following this approach, the linear term of the Cornell potential was found, with a consistent value for the string tension. Furthermore, considering the finite temperature case, the dissociation process was reproduced. These results strongly indicate that the role of the extra dimension in holographic models for hadrons can be associated with their internal structure. 
 
In this work we extend the previous one and consider the interaction energy of the quark anti-quark pair that form a charmonium state but now in a finite density medium. In Sec. \ref{sec: Holographic Model for Charmonium}, we review the tangent model presented in \cite{Braga:2017bml}, in Sec. \ref{sec: The String and the quark-antiquark potential} we summarize the holographic framework used for determining the effective quark antiquark potential in the vacuum. Then, in sec \ref{sec: finite temperature and density}, we add temperature and density by modifying the background metric. Concluding remarks and discussions are presented in section \ref{sec: Conclusions}.


\vspace{.75\baselineskip}
\section{Holographic Model for Charmonium}
\label{sec: Holographic Model for Charmonium}

In the holographic phenomenological AdS/QCD model of Ref.~\cite{Braga:2017bml}, quarkonium states are represented by a vector field $V_m$ with five components, with indices running over the coordinates $t$, $x^1$, $x^2$, $x^3$, and $z$, where the last one is the holographic coordinate. The field is described by an action integral of the form
\begin{gather}
    I = - \dfrac{1}{4 g_5^2} \intḏ^4x \int_{0}^{z_h}\, ḏz\, √{-g}\, g^{\,mp} g^{\,nq} F_{mn} F_{pq},
    \label{eq: field action}
\end{gather}
where $F_{mn} =  ∂_m V_n - ∂_n V_m$ and the metric $g$ is given by
\begin{gather}
    ḏs² = g_{mn} ḏx^m ḏx^n = \dfrac{R²}{z²} ē^{-2ϕ(z)} \PAR[3]{\!-ḏt² + (ḏx^1)² + (ḏx^2)² + (ḏx^3)² + ḏz²}.
    \label{eq: vacuum metric}
\end{gather}

It is important to note that this is a phenomenological (bottom-up) approach, in the sense that the metric is not a solution of Einstein's equations in five dimensions for some gravitational action. In other words, the back reaction of the field $ϕ(z)$ is neglected.  

In the literature, one often finds a factor $ē^{-ϕ(z)}$ directly in the action instead of the factor of $ē^{-2ϕ(z)}$ inside the metric. The field equations arising from these two different ways of writing the metric are the same, because the action itself is the same. However, there is an important difference when one considers the so called dilaton factor to be part of the metric. In this case this factor will affect a string that connects two heavy quarks and extends into the 5-d space. Such string represents their interaction \cite{Andreev:2006ct,Colangelo:2010pe}. 

The dilaton function $ϕ$ of Ref.~\cite{Braga:2017bml} was build with the purpose of fiting not only the masses of charmonium states, but also their decay constants. In order to fit also bottomonium masses and decay constants, a new linear term was included on Ref.~\cite{Braga:2018zlu}. None of these dilatons produce confinement when considered as part of the metric as in \eqref{eq: vacuum metric}. Confinement is obtained if one changes the sign of the quadratic term, as it was done in Ref.~\cite{Braga:2024heu} in order to describe the effective potential of charmonium states. The dilaton used there is
\begin{gather}
    ϕ(z) = -κ^2 z^2 - Mz + \tanh\!\PAR[3]{\dfrac{1}{Mz} - \dfrac{κ}{√{Γ}}},
    \label{eq: dilaton tangent}
\end{gather}
with $κ = \SI{1.1}{\giga\electronvolt}$, $M =  \SI{0.11}{\giga\electronvolt}$ and $√{Γ} = \SI{0.26}{\giga\electronvolt}$. These parameters were fixed by looking for the best fit for charmonium masses and decay constants at zero temperature, as well as for the dissociation temperature at zero chemical potential. The results for masses and decay constants are shown in Table I.

\begin{table}[htb]
\centering
\begin{tabular}{
    |m{1.2cm}<{\centering}
    |m{3.2cm}<{\centering}
    |m{3.2cm}<{\centering}
    |m{3.2cm}<{\centering}
    |m{3.2cm}<{\centering}
    |
}
    \hline
    \multicolumn{5}{|c|}{Charmonium Masses and Decay Constants} \\\hline
    State &
    Experimental Masses ($\si{\mega\electronvolt}$) &
    Masses on the tangent model ($\si{\mega\electronvolt}$) &
    Experimental Decay Constants ($\si{\mega\electronvolt}$) &
    Decay Constants on the tangent model ($\si{\mega\electronvolt}$)
    \\\hline
    $1S$  & $ 3096.900    \pm 0.006    $ & $2399$  & $ 416 \pm 4 $ & 298 \\\hline
    $2S$  & $ 3686.097    \pm 0.011    $ & $3560$  & $ 294.3    \pm 2.5    $ & 258 \\\hline
    $3S$  & $ 4040 \pm 4 $ & $4011$  & $ 187 \pm 8 $ & 239 \\\hline
    $4S$  & $ 4415 \pm 5 $ & $4590$  & $ 161 \pm    10 $ & 229 \\\hline
\end{tabular}\\[10pt]
\caption{Charmonium masses and decay constants. experimental data and results from the tangent model. }
\label{tab: charmonium masses and decay constants}
\end{table}


\vspace{.75\baselineskip}
\section{The String and the quark-antiquark potential}
\label{sec: The String and the quark-antiquark potential}

In the context of the gauge-gravity duality, the interaction between a quark  antiquark pair of infinite masses is represented by a string connecting them. The string endpoints, corresponding to the quark positions, are located at $z = 0$, $ x \equiv x^1 = \pm r/2 $.  The string penetrates into the holographic coordinate $z$ of the $\AdS_5$ space and has a shape that minimizes the area of the corresponding worldsheet \cite{Maldacena:1998im, Kinar:1998vq}. The Nambu-Goto action for this static string has the following form:
\begin{align}
    \SNG &= \dfrac{\Delta t}{2πα'} \int ḏx √{-g_{tt} g_{xx} - g_{tt} g_{zz} (z')^2},
    \label{eq: nambu-goto action}
\end{align}
where $α'$ is a constant with mass dimension $-2$ and $\Delta t$ is the time interval.  

\begin{figure}
    \centering
    \includegraphics[width=0.6\linewidth]{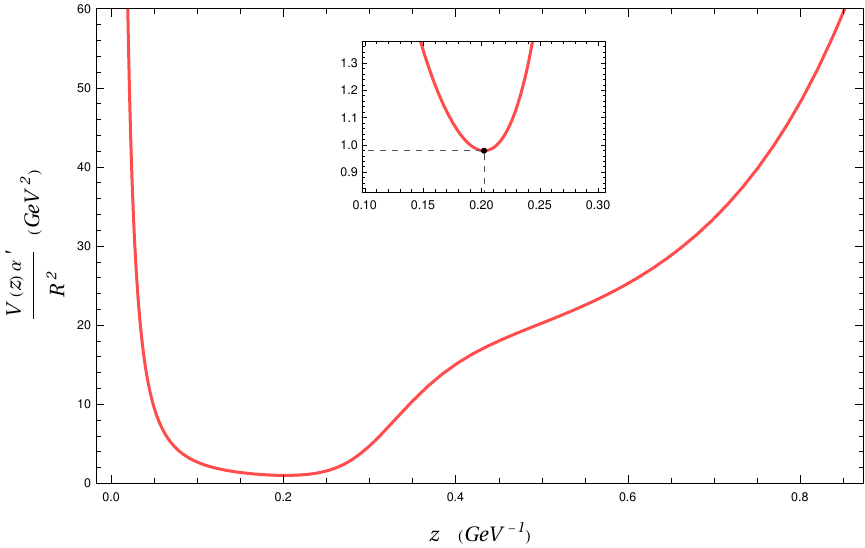}
    \caption{$V(z) α'/ R^2$ as a function of the coordinate $z$}
    \label{fig: plot of V}
\end{figure}

\begin{figure}[ht]
    \centering
    \begin{subfigure}[t]{.4\linewidth}
        \includegraphics[width=\linewidth]{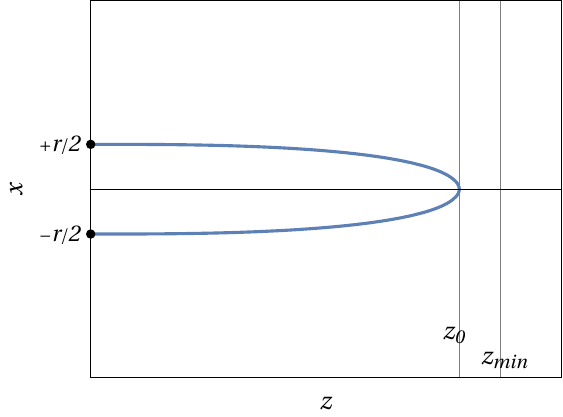}
        \caption{$ z_0 = 0.9\zmin $}
    \end{subfigure}
    \hspace{3em}
    \begin{subfigure}[t]{.4\linewidth}
        \includegraphics[width=\linewidth]{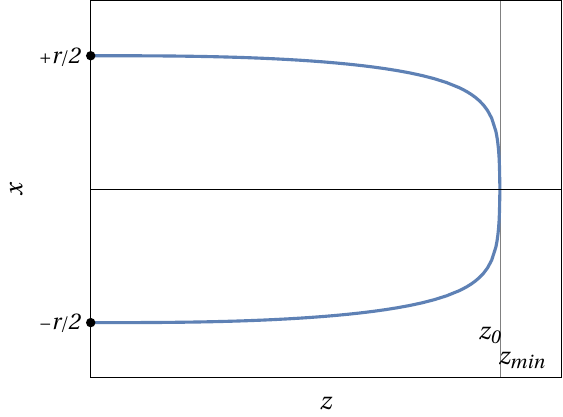}
        \caption{$ z_0 = 0.999\zmin $}
    \end{subfigure}
    \caption{Representation of two strings that are solutions of \eqref{eq: geodesic} with different values of $z_0$, corresponding to different distances $r$ between quarks.}
    \label{fig: geodesic}
\end{figure}

Following Ref.~\cite{Kinar:1998vq} one can define
\begin{align}
    V(z) &\equiv  \dfrac{1}{2πα'} \sqrt{ - g_{tt} g_{xx} }= \dfrac{1}{2πα'} \dfrac{R^2}{z^2} ē^{-2ϕ(z)}
    \label{eq: V at T=0}
\shortintertext{and}
    W(z) &\equiv \dfrac{1}{2πα'} \sqrt{ - g_{tt} g_{zz}}  = \dfrac{1}{2πα'} \dfrac{R^2}{z^2} ē^{-2ϕ(z)} = V(z).
    \label{eq: W at T=0}
\end{align}
The function $V(z)$ is plotted in Fig.~\ref{fig: plot of V}. We will call $\zmin$ the point where the minimum of this function occurs. The extremization of the Nambu-Goto action \eqref{eq: nambu-goto action} gives the equation that defines the shape of the string:
\begin{gather}
    z' = +- \dfrac{V}{W} √{V^2/V_0^2 - 1},
    \label{eq: geodesic}
\end{gather}
where the prime stands for the derivative with respect to $x$; the plus and minus signs refer to the $x<0$ and $x>0$ sides of the string, respectively; $ V_0 = V(z_0) $ and $z_0$ is the value of $z$ where the string crosses the $z$ axis, so that $z_0 = z(x=0)$. This point depends on the distance between the quarks. Two examples of this type of string, for different values of $z_0$, are shown in Fig.~\ref{fig: geodesic}.

The coordinate distance $r$ between the quarks can be written in terms of $z_0$ as \cite{Kinar:1998vq, Braga:2024heu}
\begin{gather}
    r = 2\int_{0}^{r/2} ḏx = 2\int_{0}^{z_0} \frac{ḏz}{z'} = 2 \int_{0}^{z_0} \dfrac{1}{√{V^2/V_0^2 - 1}} \dfrac{W}{V}\, ḏz.
    \label{eq: dist as function of z0}
\end{gather}
The string energy $E$ is equal to the Nambu-Goto action \eqref{eq: nambu-goto action} divided by the time interval. It is infinite, since it includes the infinite quark masses, but it can be renormalized by subtracting a factor associated with two horizontal lines with fixed points on the position of the quarks at the boundary ($z=0$), and that stretch along the holographic coordinate. Then one writes the energy in terms of $z_0$ as
\begin{gather}
    E = 2 \int_{0}^{z_0^{}} \PAR[4]{\dfrac{1}{√{1 - V_0^2/V^2}} - 1} W\, ḏz - 2 \int_{z_0^{}}^{\zmin} W ḏz.
    \label{eq: regularized energy as function of z0}
\end{gather}
As explained in \cite{Braga:2024heu}, the values of $z_0$ are limited to the interval $(0, \zmin)$, where $\zmin$ is the point of minimum of the function $V(z)$ (see Fig.~\ref{fig: plot of V}).

 One cannot write a general explicit expression for $E(r)$. However, it is simple to plot $E$ as function of $r$ using Eqs.~\eqref{eq: dist as function of z0} and \eqref{eq: regularized energy as function of z0} with $z_0$ as parameter. Also, in the limit  $z_0 --> \zmin$, where  both, $r$ and $E$, diverge with $\ln(1 - z_0/\zmin)$ \cite{Kinar:1998vq, Braga:2024heu}, one finds
\begin{flalign}
    && E(r) = V(\zmin)\, r
    && \llap{(large $r$). \hspace{1.8em}}
    \label{eq: E for large r}
\end{flalign}

This model, therefore, reproduces the confinement property of  strong interaction, in the form of the Cornell potential for large distances. By comparing Eq.~\eqref{eq: E for large r} with the Cornell potential,
\begin{gather}
    E(r) = -\frac{4}{3}\frac{α_s}{r} + σ r,
    \label{eq: Cornell potential}
\end{gather}
in the large distance limit, we find the string tension:
\begin{gather}
    σ = V(\zmin) = \dfrac{1}{2πα'} \dfrac{R^2}{\zmin^2} ē^{-2ϕ(\zmin)}
    \label{eq: string tension}
\end{gather}
In Ref.~\cite{Braga:2024heu}, the constant $R^2/\alpha'$ was fixed by solving a Schrodinger equation with the parametrized effective potential found on \eqref{eq: dist as function of z0} and \eqref{eq: regularized energy as function of z0}, and fitting the masses of charmonium states. This way, the string tension obtained was
\begin{gather}
    σ = \SI{0.17}{\giga\electronvolt^2}  ,
\end{gather}
in agreement with the results found, for example, in \cite{Ebert:2011jc, Soni:2017wvy}:   $σ = \SI{0.18}{\giga\electronvolt^2}$ and in \cite{ Mateu:2018zym}: $ σ =  \SI{0.164}{\giga\electronvolt^2}$.


\vspace{.75\baselineskip}
\section{Finite Temperature and Density}
\label{sec: finite temperature and density}

Finite temperature and density are introduced by changing the geometry to an $\AdS_5$ Reissner–Nordström metric \cite{Colangelo:2010pe,Colangelo:2012jy,Colangelo:2011sr,Lee:2009bya}, which is a solution of Einstein-Maxwell's equations, with a gauge field of the form $A_m = (A_t(z), 0, 0, 0, 0)$, with $A_t(z) = μ + k q z^2$, where $q$ is the black hole charge. The constant $k$ is related to the 5-d Newton's constant $G_5$ and to the constant $g_5$ associated with the field $A_t$. The relation comes from the solution of the Einstein's equations,
which give $k = \sqrt{\frac{3}{2} \frac{g_5^2 R^2}{8 π G_5}}$. The constants $G_5$ and $g_5$ can be written in terms of the AdS radius $R$ \cite{Veschgini:2010ws}, giving $k = \sqrt{4 N_c/5}$, where $N_c = 3$ is the number of colors. Thus: $k = \sqrt{12/5}$. The boundary value of the field $A_t$ is the chemical potential $μ$ of the plasma.

Then, for modeling the interaction of the plasma with the heavy meson, we introduce the dilaton into the metric, as in \eqref{eq: vacuum metric}. The metric, then, becomes
\begin{gather}
    ḏs² = g_{mn} ḏx^m ḏx^n = \dfrac{R²}{z²} ē^{-2 \phi(z)} \PAR[3]{\!-f(z) ḏt² + ḏ\xx² + \frac{1}{f(z)}ḏz²}.
    \label{eq: metric with black hole and dilaton}
\shortintertext{with}
    f(z) = 1 - \PAR[2]{ 1 + q^2 z_h^6 } \frac{z^4}{z_h^4} + q^2 z^6, \label{eq: bh factor}
\end{gather}
where $z_h$ is the the black hole horizon. Note that the field $A_t$ and the the metric \eqref{eq: metric with black hole and dilaton} solve Einstein-Maxwell equations only if we set $\phi = 0$.  

The temperature of the plasma is identified with the Hawking temperature of the black hole, given by
\begin{gather}
    T = \frac{1}{\pi z_h} \PAR{ 1 - \frac{1}{2} q^2 z_h^6 }.
    \label{eq: temperature function of the charge}
\end{gather}
The relation between the chemical potential and the BH charge is obtained by the condition $A_t(z_h) = 0$, which gives
\begin{gather}
    μ = k q z_h^2,
\end{gather}

Note that $q$ and $z_h$ are quantities related to the geometry of the space, while $T$ and $μ$ represent the physical properties of the plasma.



Now, using the same definition of Eq.~\eqref{eq: V at T=0} and the metric  \eqref{eq: metric with black hole and dilaton}, one finds 
\begin{equation}
    V(T, μ, z) =  \dfrac{1}{2πα'} \sqrt{ - g_{tt} g_{xx}} = \dfrac{1}{2πα'} \dfrac{R^2}{z^2} ē^{-2ϕ(z)}\sqrt{f(z)},
    \label{eq: V at finite T}
\end{equation}
while, from \eqref{eq: W at T=0}, one finds that $W(z)$ does not change. Note, considering Eq.~\eqref{eq: bh factor}, that  \(V\)  is as a function of temperature and chemical potential, through $z_h$ and $q$. 

\begin{figure}[ht]
    \centering
    \includegraphics[width=0.4\linewidth]{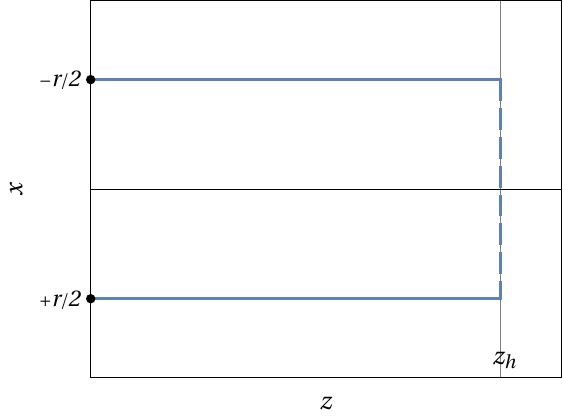}
    \qquad
    \includegraphics[width=0.4\linewidth]{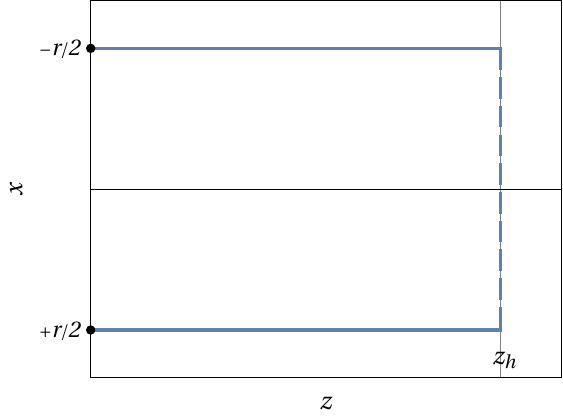}
    \caption{String configurations, for two different distances, corresponding to free quarks. Or, equivalently, to the dissociation of the meson.}
    \label{fig: deconfined representation}
\end{figure}

\begin{figure}[ht]
    \centering
    \begin{subfigure}[t]{.45\linewidth}
        \includegraphics[width=\linewidth]{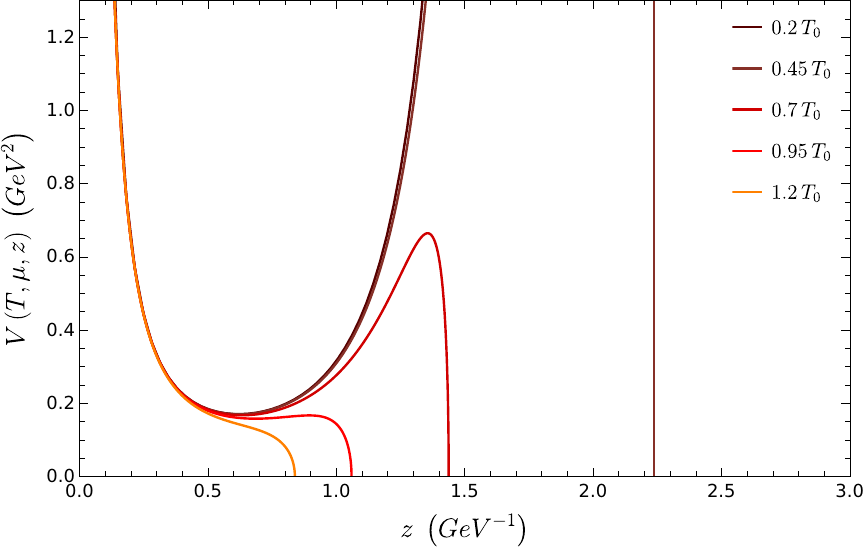}
        \caption{\(μ=0.0\)}
        \label{fig: V at zero mu}
    \end{subfigure}\qquad
    \begin{subfigure}[t]{.45\linewidth}
        \includegraphics[width=\linewidth]{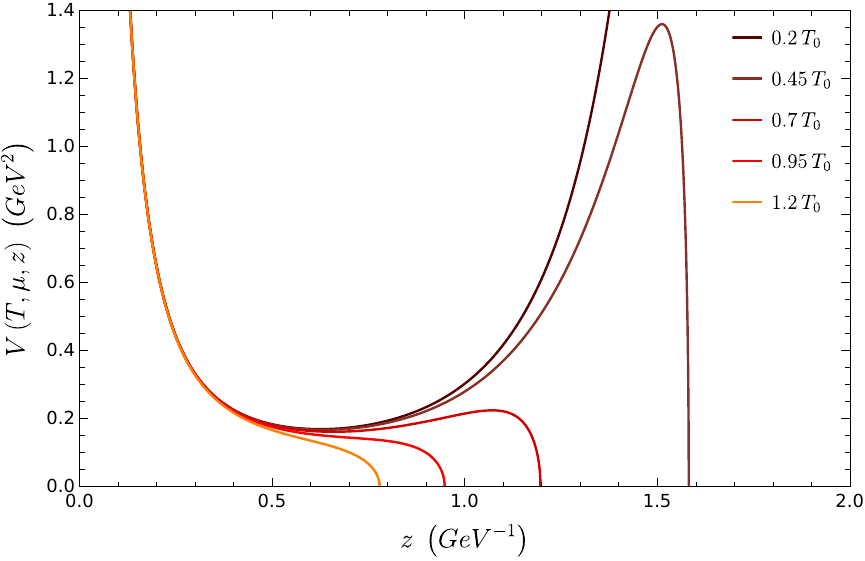}
        \caption{\(μ=0.4μ_0\)}
        \label{fig: V at .4 mu}
    \end{subfigure}\\[\baselineskip]
    \begin{subfigure}[t]{.45\linewidth}
        \includegraphics[width=\linewidth]{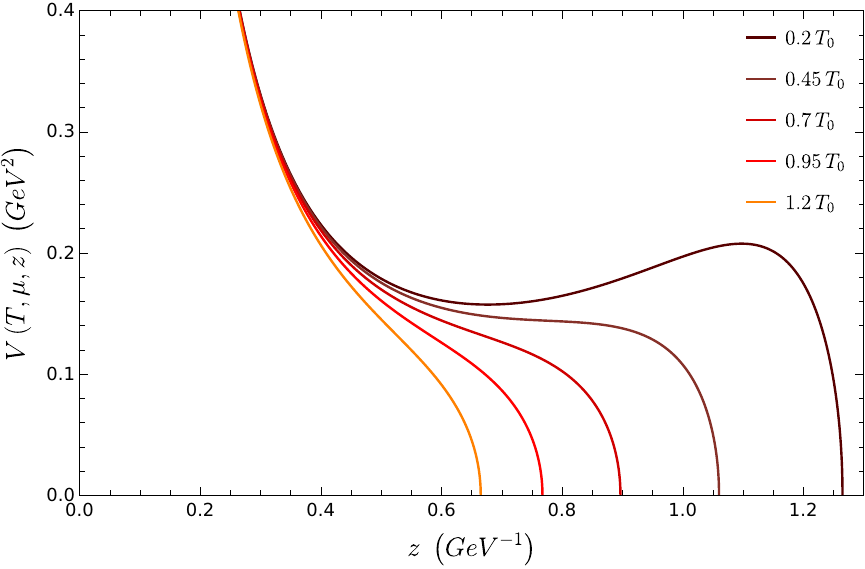}
        \caption{\(μ=0.8μ_0\)}
        \label{fig: V at .8 mu}
    \end{subfigure}\qquad
    \begin{subfigure}[t]{.45\linewidth}
        \includegraphics[width=\linewidth]{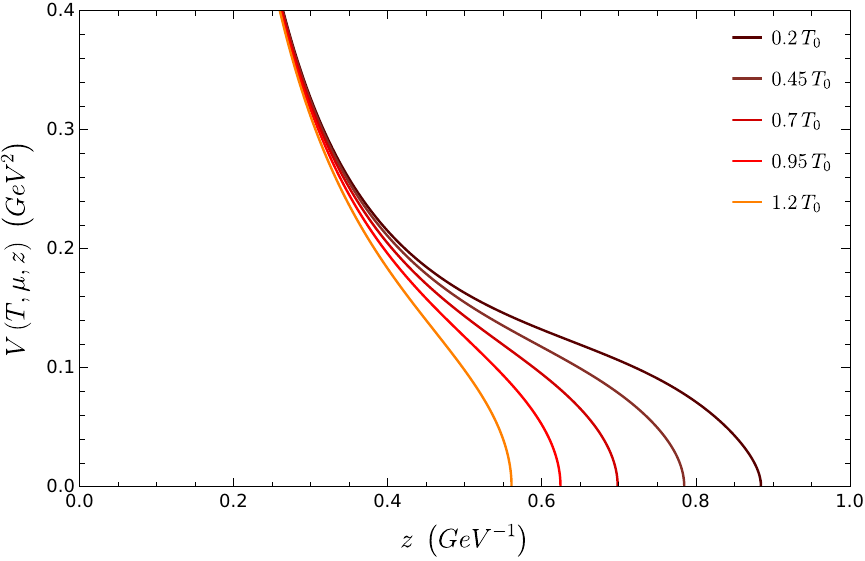}
        \caption{\(μ=1.2μ_0\)}
        \label{fig: V at 1.2 mu}
    \end{subfigure}
    \caption{ \(V\) as a function of \(z\) for various temperatures and chemical potentials}
    \label{fig: V with temperatures and desinties}
\end{figure}

By analyzing how \( V(T, μ, z) \) is affected by \(T\) and \(μ\), one can understand the dissociation process. At low temperatures and/or chemical potentials, \( V(T, μ, z) \) exhibits a non vanishing local minimum at some \( \zmin(T, μ) \), in addition to a global null minimum at \( z_h(T, μ) \), due to the presence of \( f(z) \) in \eqref{eq: V at finite T}. As a result, a local maximum emerges at some point between \( \zmin(T, μ) \) and \( z_h(T, μ) \). The existence of a non vanishing \( V(T, μ, \zmin) \) implies a non-zero string tension, which indicates confinement.

On the other hand, as the temperature and/or the chemical potential increase, the local minimum turns into an inflection point.  Consequently, the only remaining minimum is at \( z_h(T, μ) \), where \( V(T, μ, z_h) = 0 \), implying a null string tension and, therefore, the absence of confinement. Furthermore, from Eq.~\eqref{eq: geodesic}, one can note that the only possible string configuration in this situation is two straight lines extending from $0$ to $z_h$, representing two free quarks, as shown in Fig.~\ref{fig: deconfined representation}.

Therefore, temperature and chemical potential of dissociation are defined as the sets of values $\{\bar{T}, \bar{μ}\}$ in which the local minimum of $V(T, μ, z)$ becomes an inflection point. In \cite{Braga:2024heu}, this dissociation process is described in detail for the case $μ=0$, where the dissociation temperature $T_0 = \SI{316}{\mega\electronvolt}$ was found.

Now, considering the case where \(T = 0\) and using the same procedure, one finds the critical dissociation chemical potential \(μ_0 = \SI{1.7661}{\giga\electronvolt}\) predicted by this model. Since it is difficult to find experimental data for this critical quantity for charmonium, it is useful to compare this result with one obtained using a model for light mesons. For example, in \cite{DeWolfe:2010he}, the ratio $ μ_0/T_0$ found is \(5.475\), while here we obtain \(5.585\).  So, the ratio is similar, while it is important to note that each of the critical quantities, $μ_0$ and $T_0$ have different values for light and heavy mesons.

\begin{figure}[ht]
    \begin{subfigure}[h]{0.48\textwidth}
        \includegraphics[width=1\linewidth]{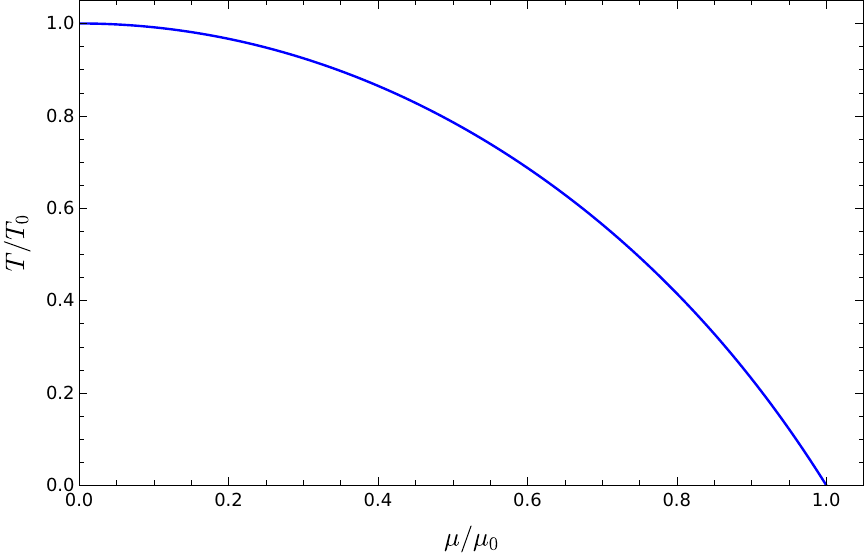}
        \caption{}
        \label{fig: dissociation curve}
    \end{subfigure}
    \hfill
    \begin{subfigure}[h]{0.48\textwidth}
        \includegraphics[width=1\linewidth]{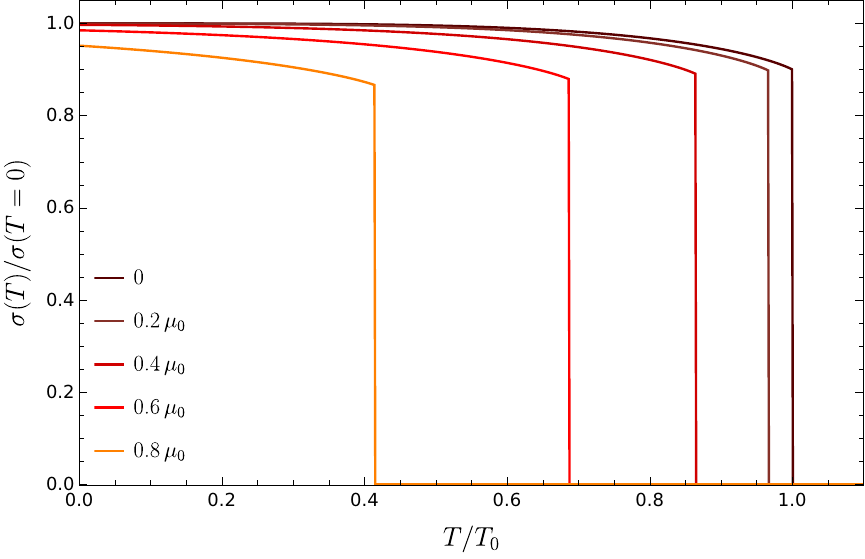}
        \caption{}
        \label{fig: String tension as function temperature}
    \end{subfigure}
    \caption{(a) Dissociation curve for charmonium. (b) String tension as a function of temperature for various chemicals potentials}
\end{figure}

For a given \(\bar{μ}\), one can numerically determine the temperature \(\bar{T}\) by calculating the first and second derivative of \(V(T, \bar{μ}, \zmin)\) and looking for the temperature where both vanish. This is the condition used in order to identify critical points in the system. By repeating this process for all values of \(\bar{μ}\)   in the interval \([0, μ_0]\), one obtains the plot shown in Fig.~\ref{fig: dissociation curve}, which represents the critical curve for charmonium. In this plot, quarks are confined (bound states) below the blue line and deconfined (free) above it.


In this system, the string tension $\sigma(T, μ)$ is the order parameter for the confined{\slash}deconfined phase transition. The dependence of $\sigma(T, μ)$ with the temperature is plotted in Fig.~\ref{fig: String tension as function temperature} for various chemical potentials. The key observation from this figure is that at the dissociation temperatures $\bar T$ the string tension jumps from a finite value to zero, remaining zero for all higher temperatures. This jump in the order parameter characterizes a first-order transition and produces a singular point in the free energy of the system at $\bar T$, as we show in the next section. 

\subsection{Quark anti-quark  energy }

\begin{figure}[ht]
    \begin{subfigure}[h]{0.48\textwidth}
        \includegraphics[width=1\linewidth]{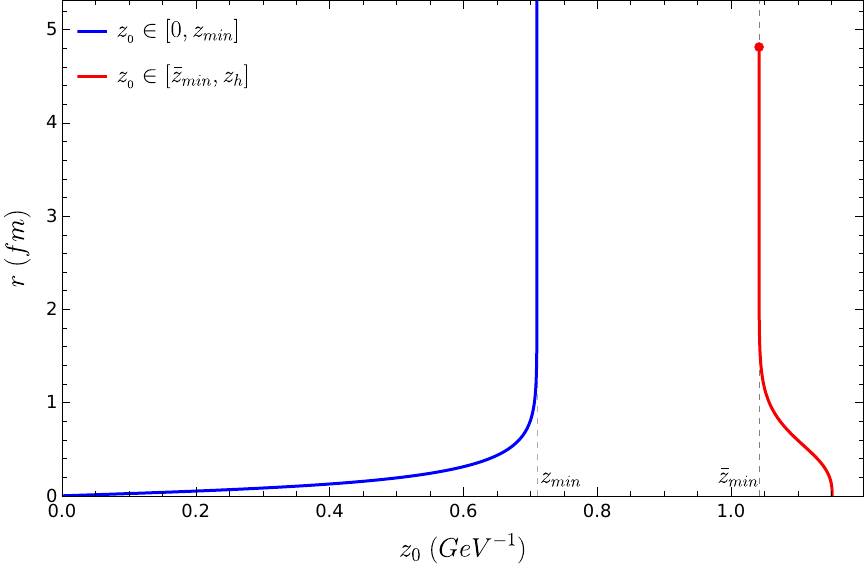}
        \caption{}
        \label{fig: distance as function of z0 below T bar}
    \end{subfigure}
    \hfill
    \begin{subfigure}[h]{0.48\textwidth}
        \includegraphics[width=1\linewidth]{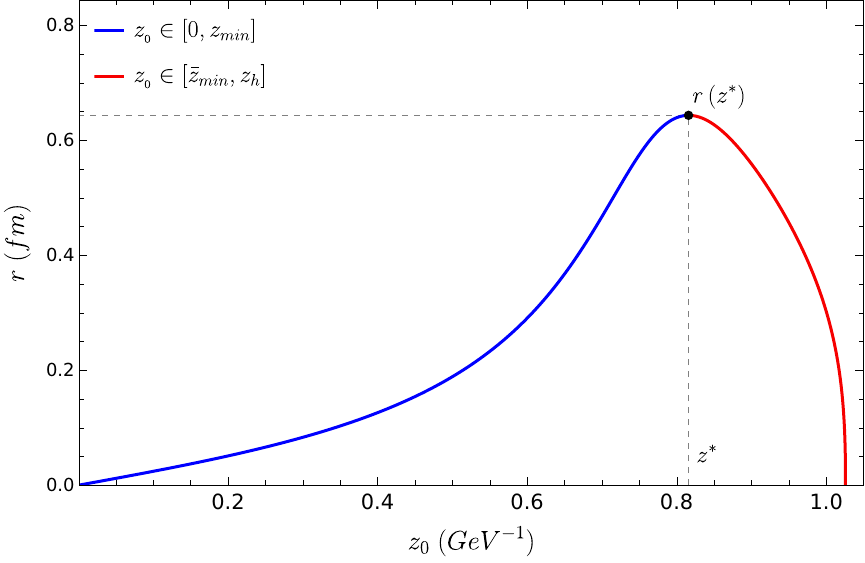}
        \caption{}
        \label{fig: distance as function of z0 above T bar}
    \end{subfigure}
    \caption{Distance of dissociation in the cases of (a) below \(\{\bar{T}, \bar{μ}\}\), (b) above \(\{\bar{T}, \bar{μ}\}\)}
\end{figure}

Let us analyze how the interaction energy, which in this context corresponds to the grand potential \(\Phi \), depends on the quark anti-quark distance $r$ as the temperature and the chemical potential vary. From the discussion in the previous section, it is not possible to find an explicit relation of the form  \(\Phi(T, μ, r)\). The quark distance and the energy are expressed in terms of the quantity $z_0$ in Eqs.~\eqref{eq: dist as function of z0} and \eqref{eq: regularized energy as function of z0} respectively, now using $V$ from Eq.~\eqref{eq: V at finite T}.
So, let us start considering the variation of $r$ in terms of $z_0$. 
For temperatures and chemical potentials in the region below the critical curve of Fig.~\ref{fig: dissociation curve}, that corresponds to the confined phase, one can use Eq.~\eqref{eq: dist as function of z0} to calculate the separation distance for \(z_0 \in[0, \zmin(T, μ)]\). The result is similar to the zero temperature case: the distance increases with \(z_0\) and diverges at \(\zmin(T, μ)\). This behavior is represented by the blue line in Fig.~\ref{fig: distance as function of z0 below T bar}.

From Eq.~\eqref{eq: geodesic} one can take the relation 
\(V_0(z)^2\le V(z)^2\), which imposes a condition on string formation. This condition is satisfied in the region \(z_0 \in[0, \zmin(T, μ)]\). However, unlike the zero temperature case,  a second permitted region exists in \(z_0 \in [{\zminbar}(T, μ), z_h(T, μ)]\), where \({\zminbar}(T, μ)\) is the point at which \(V({\zminbar}(T, μ))= V(\zmin(T, μ))\). Applying Eq.~\eqref{eq: dist as function of z0} to the second region, one finds that the separation distance has a finite value at \({\zminbar}(T, μ)\) and decreases to zero at \(z_h(T, μ)\), as shown by the red line in Fig.~\ref{fig: distance as function of z0 below T bar}.

For  \((T, μ) > \{\bar{T}, \bar{μ}\}\), the minimum is \(V(T, μ, z_h) =0\), so the relation \(V_0(z)^2\le V(z)^2\) is satisfied for all \(z_0\in [0, z_h(T, μ)]\). This implies that the string can extend from  0 to $z_h$. Applying \eqref{eq: dist as function of z0} once again, one finds the curve plotted in Fig.~\ref{fig: distance as function of z0 above T bar}. The separation distance increases when $z_0 $ goes from \(0\) to a value \(z^*(T, μ)\), which corresponds to the \(z\) of the inflection point, mentioned in the previous section, and then decreases to zero at \(z_h(T, μ)\).

One can generalize Eq.~\eqref{eq: regularized energy as function of z0} in order to derive the grand potential \(\Phi(T, μ, z_0)\). Before proceeding with the calculations it is important to remark that we use the same regularization used in zero temperature and chemical potential. 

\begin{figure}[ht]
    \centering
    \begin{subfigure}[t]{.45\linewidth}
        \includegraphics[width=\linewidth]{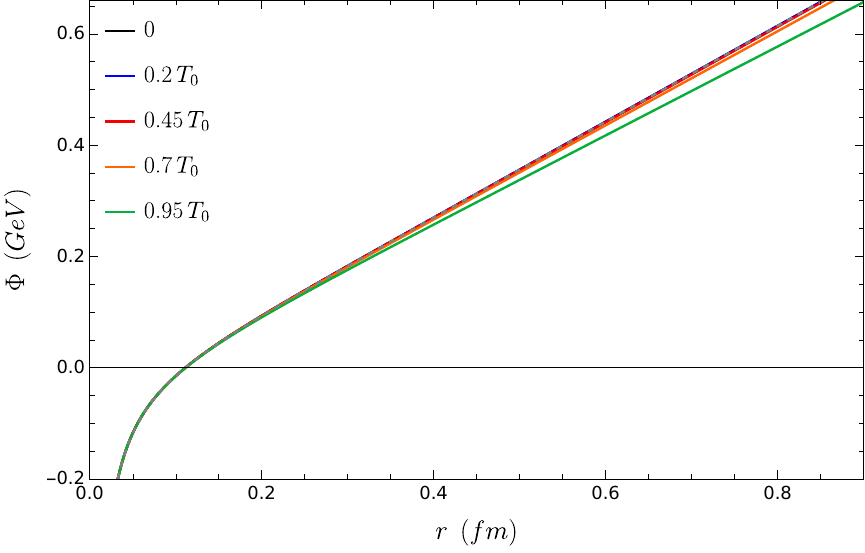}
        \caption{\(μ=0.0\)}
        \label{fig: energy at 0.0 mud}
    \end{subfigure}\qquad
    \begin{subfigure}[t]{.45\linewidth}
        \includegraphics[width=\linewidth]{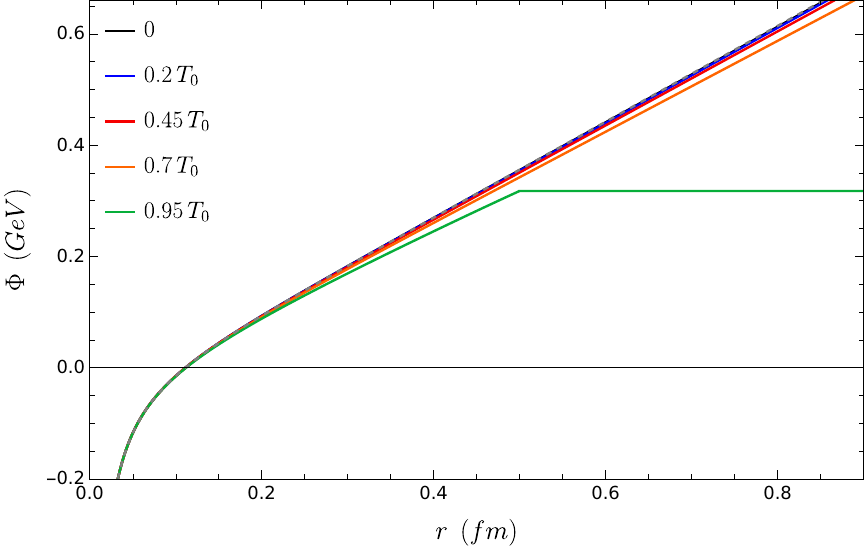}
        \caption{\(μ=0.4μ_0\)}
        \label{fig: energy at 0.4 mud}
    \end{subfigure}\\[\baselineskip]
    \begin{subfigure}[t]{.45\linewidth}
        \includegraphics[width=\linewidth]{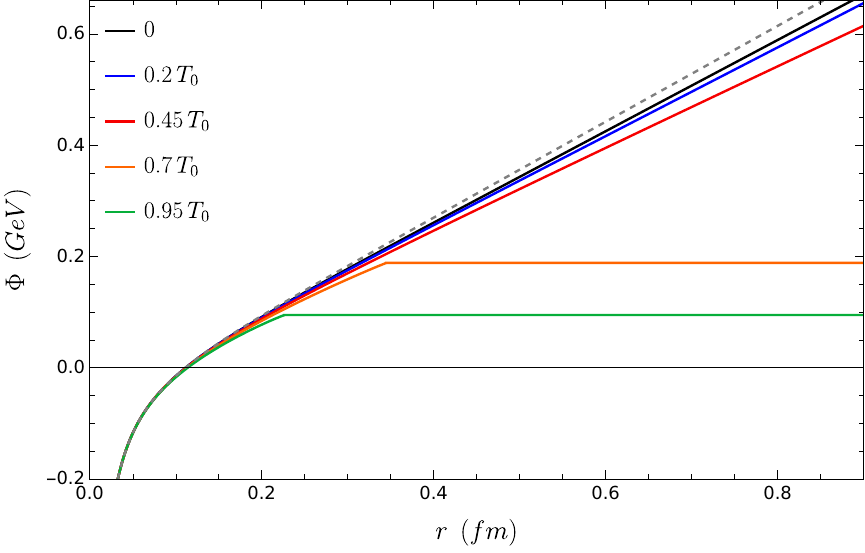}
        \caption{\(μ=0.8μ_0\)}
        \label{fig: energy at 0.6 mud}
    \end{subfigure}\qquad
    \begin{subfigure}[t]{.45\linewidth}
        \includegraphics[width=\linewidth]{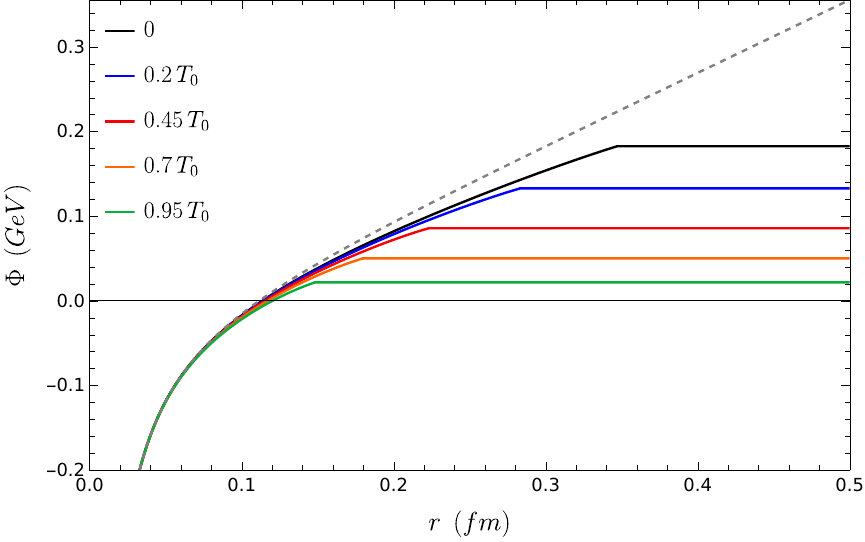}
        \caption{\(μ=1.2μ_0\)}
        \label{fig: energy at 1.2 mud}
    \end{subfigure}
    \caption{Grand canonical potential as a function of quark separation for various values of temperature and chemical potential. The dashed line shows the case $T=0$, $μ =0$ for comparison.  }
    \label{fig: energy for various mu}
\end{figure}

To get the grand potential for \((T, μ) < \{\bar{T}, \bar{μ}\}\), one can use \eqref{eq: regularized energy as function of z0} in both permitted regions. However, the region \(z_0 \in [\zminbar(T, μ), z_h(T, μ)]\) gives the highest value of energy in comparison to the first region for the same separation distance, then it can be dropped because the energy should be minimized. 

Calculating the grand potential in the first region and using the parametrization \((r(T, μ,z_0), \Phi(T,μ, z_0))\) one obtains a shape similar to the Cornell potential, which is characteristic of a confining potential. This potential exhibits a slight decrease in string tension as \(T\) and/or \(μ\) increases, as can be seen in Fig.~\ref{fig: energy at 0.0 mud}.

On the other hand, when calculating \(\Phi(T,μ, z_0)\) for \((T, μ) > \{\bar{T}, \bar{μ}\}\), one observes that the grand potential increases with     \(z_0 \) in the interval \( [0, z^*(T, μ)]\) and decreases for \(z_0 \in [z^*(T, μ), z_h(T, μ)]\). Using the parametrization \((r(T, μ, z_0), \Phi(T,μ, z_0))\), one finds that \(z_0 \in [z^*(T, μ), z_h(T, μ)]\) corresponds to higher energy values for the same separation distances of \(z_0 \in [0, z^*(T, μ)]\). Since higher energy values are less favored, this region will be ignored. 

Therefore, the result is a potential with a maximum range, \(r(T, μ, z^*_0)\). For separation distances greater than  \(r(T, μ, z_0^*)\), the quarks are free. To determine the energy of these free quarks, one must calculate the energy of the two straight lines from \(0\) to \(z_h(T, μ)\). As the lines have no dependence in the \(x\) coordinate, the energy becomes 
\begin{align}
    \Phi_{\textrm{free}}(T,μ, z_0) &= \frac{1}{2\pi\alpha'}\int_{0}^{z_h(T, μ)}\sqrt{-g_{tt}g_{zz}} ḏz\nonumber\\
    &=\int_{\epsilon(T, μ)}^{z_h(T, μ)}W(z) ḏz.
\end{align}
note that a regularization parameter \(\epsilon(T, μ)\) has been introduced to remove the divergence at \(z=0\). This parameter is determined in each point of the set \(\{\bar{T}, \bar{μ}\}\) to ensure that the grand potential is a continuous function of the separation distance \(r\). The \(\Phi(T, μ, r)\) resulting from this construction is shown in Fig.~\ref{fig: energy at 0.4 mud}, \ref{fig: energy at 0.6 mud} and \ref{fig: energy at 1.2 mud}, where the \(μ\) effect is emphasized.

The physical system that emerges from this construction is interesting. To analyze it, first consider a string with large \(r\). As \(T\) and \(μ\) increase, starting from points $(T, μ) $ in the region below the curve of Fig. 5a,  the string tension slightly decreases. When a point $ (\bar T, \bar{μ})$ is reached, it jumps to zero and the configuration suddenly changes from the string as in the Fig.~\ref{fig: geodesic} to two straight lines, as depicted in Fig.~\ref{fig: deconfined representation}. This change in configuration represents a first-order phase transition between confined and deconfined phases. 

Now, for the case \((T, μ) > \{\bar{T}, \bar{μ}\}\), the energy  is represented by the green line in Fig.~\ref{fig: energy at 0.4 mud}, for example. For small separation distances, the quarks are subject to a Cornell-like potential and, therefore, remain in a bound state. However, for \(r > r(T, μ, z_0^*)\), the quarks become free, with a constant energy given by the two-line configuration. This indicates that \(r(T, μ, z_0^*)\) is the dissociation distance at which the confining-deconfining transition occurs.

\section{Conclusions}
\label{sec: Conclusions}

The holographic tangent model proposed in Refs. \cite{Braga:2017bml, Braga:2018zlu}
provides a phenomenological description of quarkonium spectra of masses and decay constants in the vacuum and also of the thermal behavior when a medium like the QGP is present. As it is the case in general for holographic models, the hadronic states/quasi-states are represented by fields living in a five dimensional space. 

It was recently shown in Ref.~\cite{Braga:2024heu}, for the case of charmonium at zero chemical potential, that the 5-d background of the tangent model can be associated with the meson's internal structure, composed of a charm anti-charm quark pair. The strong interaction between the quarks is represented by a string with endpoints fixed at the position of the quarks, and with a form that penetrates into the holographic coordinate. The renormalized energy of this string is interpreted as the interaction energy between the quarks. At zero temperature, this description gives a potential which, as the Cornell Potential, grows linearly with the distance $r$ between the quarks in the large $r$ limit. This property characterizes linear confinement. At finite temperature the dissociation occurs at some critical value of $T$ where the string tension vanishes.

In the present work, we extended the proposal of Ref.~\cite{Braga:2024heu} to the case of a medium with finite chemical potential $μ$, i.e. finite density. It was shown that increasing $μ$ has an effect similar to increasing $T$: it enhances the dissociation process. One of the main results found is the dissociation curve, or phase diagram, shown in Fig.~\ref{fig: dissociation curve}, that represents the dissociation temperature as a function of the chemical potential. It was also show that when $T$ and/or $μ$ increase, the string tension decreases, going to zero abruptly at the critical points $\{\bar T, \bar{μ} \}$ shown in Fig.~\ref{fig: String tension as function temperature}. So, it was shown here that the interpretation of the holographic coordinate as representing the interaction between the quarks holds also for the finite chemical potential case.


\hspace{\baselineskip}

 \noindent {\bf Acknowledgments:} The authors are partially supported by CNPq --- Conselho Nacional de Desenvolvimento Científico e Tecnológico, by FAPERJ --- Fundação Carlos Chagas Filho de Amparo à Pesquisa do Estado do Rio de Janeiro and by  Coordenação de Aperfeiçoamento de Pessoal de Nível Superior --- Brasil (CAPES), Finance Code 001.


\setstretch{1.19}
\bibliographystyle{apsrev4-2}
\bibliography{bibliography}

\end{document}